\documentclass[aps,prc,superscriptaddress,twocolumn]{revtex4}

\usepackage{amssymb}
\usepackage{graphicx}
\usepackage{epsfig}%
\usepackage{bm}
\usepackage{hyperref}
\usepackage{mathrsfs}
\usepackage{multirow}
\usepackage{footmisc}
\usepackage{cleveref}

\begin{document}

\title{$\Delta$-mass dependence of the M-matrix in the calculation of $\mbox{N}\Delta\to \mbox{NN}$ cross sections}


\author{Ying Cui}
\email{yingcuid@163.com}
\affiliation{China Institute of Atomic Energy, Beijing 102413, China}

\author{Yingxun Zhang}
\email{zhyx@ciae.ac.cn}
\affiliation{China Institute of Atomic Energy, Beijing 102413, China}
\affiliation{Guangxi Key Laboratory Breeding Base of Nuclear Physics and Technology, Guangxi Normal University, Guilin 541004, China}

\author{Zhuxia Li}
\affiliation{China Institute of Atomic Energy, Beijing 102413, China}

\begin{abstract}
Within the one boson exchange model, $\Delta$-mass dependent M-matrix and its influence on the calculation of $N\Delta \to NN$ cross sections are investigated. Our calculations show that the $m_{\Delta}$ dependence of $|\textbf{p}_{N\Delta}|$ and $|\mathcal{M}|^2$ has effects on the calculations of $\sigma_{N\Delta\to NN}$, especially around the threshold energy. We finally provide a table of accurate $\sigma_{N\Delta\to NN}$ which can be used in the transport models.
\end{abstract}

\date{\today}


\pacs{Valid PACS appear here} 

\maketitle

The production and absorption for $\Delta$ resonance in heavy ion collision around its threshold energy has attracted a lot of attentions again in recent years, because the ratios of charged pions which are decayed from $\Delta$ resonance was supposed to be a sensitive observable to probe the symmetry energy at suprasaturation density \cite{BALi02,Xiao2009,Feng2010,Xie2013}. In last ten years, different conclusions on the constraints of symmetry energy had been obtained based on the different transport models \cite{Xiao2009,Feng2010,Xie2013,Hong2014,Song2015,Cozma2017}, the situations stimulate further study to understand the $\Delta$ production and absorption mechanism as well as its sensitive density region probed by $\pi^-/\pi^+$ ratios. Very recently, Gao-Chan Yong \cite{Yong2019} claimed that $\pi^{-}/\pi^{+}$ ratio is sensitive to the symmetry energy around normal density rather than that at suprasaturation density based on the IBUU calculations. The debates on the constraints of symmetry energy at suprasaturation density by using $\pi^-/\pi^+$ ratios indicate a more careful study of the $\Delta$ production and absorption cross sections as well as the propagation of $\pi$ in the reaction is an urgent need.

Generally, in heavy ion collision at intermediate energies, the production and propagation of a pion experience following process, 1) First $\Delta$ production through $NN\to N\Delta$ collisions; 2) after about 2 fm/c which depends on the width of $\Delta$-resonances, $\Delta$s decay into nucleon and pion, and following them, $\pi$s are absorbed through $\pi+N\to \Delta$ process; 3) $\Delta$s with longer lifetime and higher energy participate the $N\Delta\to NN$ process. 
The possibility of three processes are directly related to their cross section or decay width in transport model simulations. Due to the complication of high-dimension transport models, most of the transport models adopt Monte-Carlo cascade method to solve the collision part where the nucleon-nucleon cross section and decay width are the key inputs.
For process 1) and 2), the cross sections and decay width can be measured in experiments, and there is less ambiguous. But, the cross section of $N\Delta \rightarrow NN$ in process 3) can not be measured directly in experiment, one has to calculate based on the detailed balance relationship.


One of the popular way to obtain the $N\Delta\to NN$ cross sections is to calculate it from the measured cross section of $NN\to N\Delta$ based on the detailed balance \cite{Bertsch88,Danielewicz1991,Wolf1992,Wolf1993,Engel1994,Baoanli1993,Bass98,ZhenZhang2017}, where the cross section of $NN\to N\Delta$ in free space has been measured by \cite{Baldini1987,Bugg1964,Cern8301,Cern8401,KEK} and it can be well explained with the one boson exchange model(OBEM) and relativistic Boltzmann-Uhling-Uhlenbeck approach \cite{Huber1994,Larionov2003,Mao1994prc,QingfengLi2017,Cui2018}. The detailed balance means the equality of scattering matrix elements which are obtained from the time reversal invariance, i.e. $|\mathcal{M}|_{if}^2$=$|\mathcal{M}|_{fi}^2$, $i$ and $f$ are the initial and final state of scattering particles.

Since the $\Delta$ is a resonance particle with a broad mass distribution,  it leads to the different forms on the calculations of $N\Delta\to NN$ cross section \cite{Bertsch88,Danielewicz1991,Wolf1992, Baoanli1993}. For example, Danielewicz \textit{et al.} considered the $\Delta$-mass distribution in the calculation of $\sigma_{NN\to N\Delta}$ with the linearly  $m_\Delta$ dependence of $\overline{|\mathcal{M}|^2}$ (i.e. ignored the $\Delta$-mass dependence of $\overline{|\mathcal{M}_D|^2}$) \cite{Danielewicz1991} \footnote{ The definition of $|\mathcal{M}|^2$ in this work is different from that in Danielewicz's work \cite{Danielewicz1991}. For the convenience in the following discussions, we named the M-matrix from Danielewicz's work  as $\mathcal{M}_{D}$. There is a following relationship between ours and Danielewicz's, i.e. $4m_{\Delta}m_{N}^3|\mathcal{M}_{D}|^2=|\mathcal{M}|^2$.
Thus, the mass independence of $|\mathcal{M}_{D}|^2$ means the $|\mathcal{M}|^2$ in this work should linearly increase with the mass of $\Delta$.
}in the $NN\to N\Delta$ process, and thus they obtained the following relationship of the one-$\Delta(1232)$ absorption cross section\cite{Danielewicz1991,Baoanli1993,ZhenZhang2017,ZhenZhang2018,Verwest1982,Huber1994} ,
\begin{eqnarray}
\label{eq:dbpawel}
&&\sigma_{N_{3}\Delta_{4}(m_\Delta)\to N_{1}N_{2}}=\\\nonumber 
&&\frac{1}{2}\frac{1}{1+\delta_{N_{1}N_{2}}}\frac{|\textbf{p}_{\text{12}}|^2}{|\textbf{p}_{\text{34}}(m_\Delta)|}
\sigma_{N_1N_2\to N_3\Delta}\\\nonumber
&&/{\int_{m_N+m_\pi}^{\sqrt s-m_N} dm'_\Delta f(m'_\Delta)|\textbf{p}_{\text{34}}(m'_\Delta)|}.
\end{eqnarray}
$f(m'_\Delta)$ is the $\Delta$ mass distribution, $\frac{1}{1+\delta_{N_1N_2}}$ is used for considering the identical of final two nucleons. If one also ignores the $\Delta$-mass dependence of $|\textbf{p}_{\text{34}}|$, it leads the Wolf \textit{et al}'s formula \cite{Wolf1992,Wolf1993}
\begin{eqnarray}
\label{eq:db92}
\sigma_{N_{3}\Delta_{4}\to N_{1}N_{2}}(m_\Delta)&=&\frac{1}{2N}\frac{1}{1+\delta_{N_{1}N_{2}}}\frac{|\textbf{p}_{\text{12}}|^2}{|\textbf{p}_{\text{34}}|^2} \nonumber\\
&&\times \sigma_{ N_{1}N_{2}\to N_{3}\Delta_{4}}.
\end{eqnarray}
where the factor $N=\int^{\sqrt{s}-m_N}_{m_N+m_\pi} f(m_\Delta) dm_\Delta$. The influence of both methods on the heavy ion collisions have been discussed in reference\cite{Baoanli1993,Wolf1992,Wolf1993}, and it is found that the modified form of $N\Delta\to NN$ can obviously influence the heavy ion collisions observables, such as rapidity distribution of pion and its flow, at the beam energy from 0.8 A GeV to 1.35 A GeV. Since both of these methods ignored the $\Delta$-mass dependence of $|\mathcal{M}_{D}|^2$ or $|\textbf{p}_{\text{34}}|$, which was thought to be very important near the threshold energy, it will be interesting to valuate the precision of  two methods on the calculation on the cross section of $N\Delta\to NN$ and give a $N\Delta\to NN$ cross section which consider the $m_\Delta$ dependence on M-matrix and $|\textbf{p}_{\text{34}}|$.

In this paper, we first investigate the $\Delta$-mass dependence of $\overline{|\mathcal{M}|^2}$ ($\overline{|\mathcal{M}_D|^2}$) and $|\textbf{p}_{\text{34}}(m_{\Delta})|$ within the framework of the OBEM. $\sigma_{N\Delta\to NN}$ in free space is directly obtained from M-matrix element, and it is chosen as a benchmark for checking the precision of the proposed methods\cite{Danielewicz1991,Wolf1992,Wolf1993} for calculating the $\sigma_{N\Delta\to NN}$ from $\sigma_{NN\to N\Delta}$. Finally, the precise results for $\sigma_{N\Delta\to NN}$ and the function of sampling the mass of $\Delta$ in the transport models are given. 

We adopt the OBEM method with the effective Lagrangian density for nucleon and $\Delta$ baryons interacting through
$\sigma$, $\omega$, $\rho$, $\delta$, and $\pi$ mesons\cite{Cui2018,Cui2019,Huber1994,Machleidt1987,Benmerrouche1989}. Different from the work in Ref.~\cite{Larionov2003}, we include the isovector mesons $\rho$ and $\delta$ in order to describe the isospin asymmetric nuclear matter and isospin dependent in-medium $NN\rightleftharpoons N\Delta$ cross section. Theoretically, the cross section of $NN\rightleftharpoons N\Delta$ can be calculated from the their M-matrix\cite{Huber1994}. The elementary two-body cross section of $NN\to N\Delta$ at given $m_\Delta$ reads
\begin{eqnarray}
\label{eq:xsnd2}
\tilde{\sigma}( m_{\Delta})
&=&\frac{1}{4F}\int
\frac{d^3 \textbf{p}_3}{(2\pi)^3 2E_3 }  \frac{d^3 \textbf{p}_4}{(2\pi)^3 2E_4}\\\nonumber
&&\times(2\pi)^4\delta^{4}(p_1+p_2-p_3-p_4)\overline{|\mathcal{M}|^2}\\\nonumber
&=&\frac{1}{64\pi^2}\int \frac{|\textbf{p}_{\text{34}}(m_\Delta)|}{\sqrt{s_{\text{in}}}\sqrt{s_{out}}|\textbf{p}_{\text{12}}|} \overline{|\mathcal{M}|^2}  d\Omega,
\end{eqnarray}
where $\overline{|\mathcal{M}|^2}=\frac{1}{(2s_{1}+1)(2s_2+1)}\sum\limits_{s_{1}s_{2}s_{3}s_{4}}|\mathcal{M}|^2$ is for $N_1N_2\to N_3\Delta_4$ process.
$\textbf{p}_{\text{1,2}}$ and $\textbf{p}_{\text{3,4}}(m_\Delta)$ are the center-of-mass momenta of the incoming (1 and 2) and outgoing particles (3 and 4), respectively.  $F=\sqrt{(p_{1}p_{2})^2-p^{2}_{1}p^{2}_{2}}=\sqrt{s_{\text{in}}}|\textbf{p}_{\text{in}}|$ is the invariant flux factors, $s_{\text{in}}=(p_1+p_2)^2$, and $s_{\text{out}}=(p_3+p_4)^2$. The total cross section is the elementary two-body cross section averaged over the mass distribution of $\Delta$, i.e.,
\begin{eqnarray}
\label{eq:xsnnndM}
&&\sigma_{N_{1}N_{2}\to N_{3}\Delta_{4}}\\\nonumber
&&=\frac{1}{64\pi^2} \int dm'_{\Delta} d\Omega \frac{|\textbf{p}_{\text{34}}(m'_\Delta)|}{s|\textbf{p}_{\text{12}}|} \overline{|\mathcal{M}|^2}f(m'_{\Delta})\\\nonumber
\end{eqnarray}
$f(m_{\Delta})$ is the mass distribution of $\Delta$ resonance,
\begin{equation}
\label{eq:bt}
f(m_{\Delta})=\frac{2}{\pi}\frac{m^{2}_{\Delta}\Gamma(m_{\Delta})}{(m^{2}_{0,\Delta}-m^{2}_{\Delta})^2+m^{2}_{\Delta}\Gamma^2(m_{\Delta}) }.
\end{equation}
Here, $m_{0,\Delta}$ is the pole mass of $\Delta$. The decay width $\Gamma(m_\Delta)$ is taken as a parameteric form \cite{Larionov2003}.

For $\sigma_{N_{3}\Delta_{4}\to N_{1}N_{2}}$ at the given value of $m_\Delta$ can be exactly calculated as,
\begin{eqnarray}
\label{eq:xsndnnM}
&&\sigma_{N_{3}\Delta_{4}(m_\Delta)\to N_{1}N_{2}}\\\nonumber
&=&\frac{1}{4F}\int
\frac{d^3 \textbf{p}'_2}{(2\pi)^3 2E_2}  \frac{d^3 \textbf{p}'_1}{(2\pi)^3 2E_1 }\\\nonumber
&&\times(2\pi)^4\delta^{4}(p_1+p_2-p_3-p_4)\overline{|\mathcal{M}_{N\Delta(m_\Delta) \to NN}|^2}\\\nonumber
&=&\frac{1}{1+\delta_{N_{1}N_{2}}}\frac{1}{64\pi^2}\int \frac{|\textbf{p}'_{\text{12}}|}{\sqrt{s_{\text{34}}}\sqrt{s_{\text{12}}}|\textbf{p}'_{\text{34}}(m_\Delta)|}\\\nonumber
&&\times \overline{|\mathcal{M}_{N\Delta(m_\Delta) \to NN}|^2}  d\Omega.
\end{eqnarray}
and there is,
\begin{eqnarray}
&&\overline{|\mathcal{M}_{N_3\Delta_4(m_\Delta) \to N_1N_2}|^2}\\\nonumber
&&=\frac{(2s_{1}+1)(2s_2+1)}{(2s_{3}+1)(2s_4+1)}\overline{|\mathcal{M}_{N_1N_2\to N_3\Delta_4(m_\Delta)}|^2}
\end{eqnarray}
at the same $\Delta$ mass for both process. For convenience, we use $\overline{|\mathcal{M}(m_{\Delta})|^2}$ to represent $\overline{|\mathcal{M}_{N_1N_2\to N_3\Delta_4(m_\Delta)}|^2}$ in the following description. The ratio between Eq.(\ref{eq:xsndnnM}) and Eq.(\ref{eq:xsnnndM}) can give an exact relationship between the cross section of $NN\to N\Delta$ and $N\Delta \to NN$. Thus, $\sigma_{N\Delta\to NN}$ can be written as,
\begin{eqnarray}
\label{eq:ratio}
&&\sigma_{N\Delta(m_{\Delta})\to NN}=\frac{1}{1+\delta_{N_1N_2}}\frac{(2s_{1}+1)(2s_2+1)}{(2s_{3}+1)(2s_4+1)}\times\\\nonumber
&&\frac{\int d\Omega |p_{12}|^2\overline{|\mathcal{M}(m_{\Delta})|^2}}{\int d\Omega |p'_{34}(m_{\Delta})|\int |p_{34}(m'_{\Delta})|f(m'_{\Delta})\overline{|\mathcal{M}(m'_{\Delta})|^2}dm'_{\Delta}}\sigma_{NN\to N\Delta}
\end{eqnarray}
One should notice, $\textbf{p}'_{\text{34}}$ and $\textbf{p}'_{\text{12}}$ are the momentum of $N_3$ (or $\Delta_4$) and $N_1$ (or $N_2$) in center of mass of colliding particles in the process of $N\Delta\to NN$, while $\textbf{p}_{\text{12}}$ and $\textbf{p}_{\text{34}}$ are the momentum of $N_1$ (or $N_2$)and $N_3$ (or $\Delta_4$) in the process of $NN\to N\Delta$. At given center of mass energy $\sqrt s$, there is $|\textbf{p}_{\text{12}}|$=$|\textbf{p}'_{\text{12}}|$ for ingoing nucleons and outgoing nucleons, but $|\textbf{p}_{\text{34}}|$ may not equal to $|\textbf{p}'_{\text{34}}|$ which depends on the equality of mass of $\Delta$ in its production and absorption process.

\begin{figure}[htbp]
\begin{center}
    \includegraphics[scale=0.31]{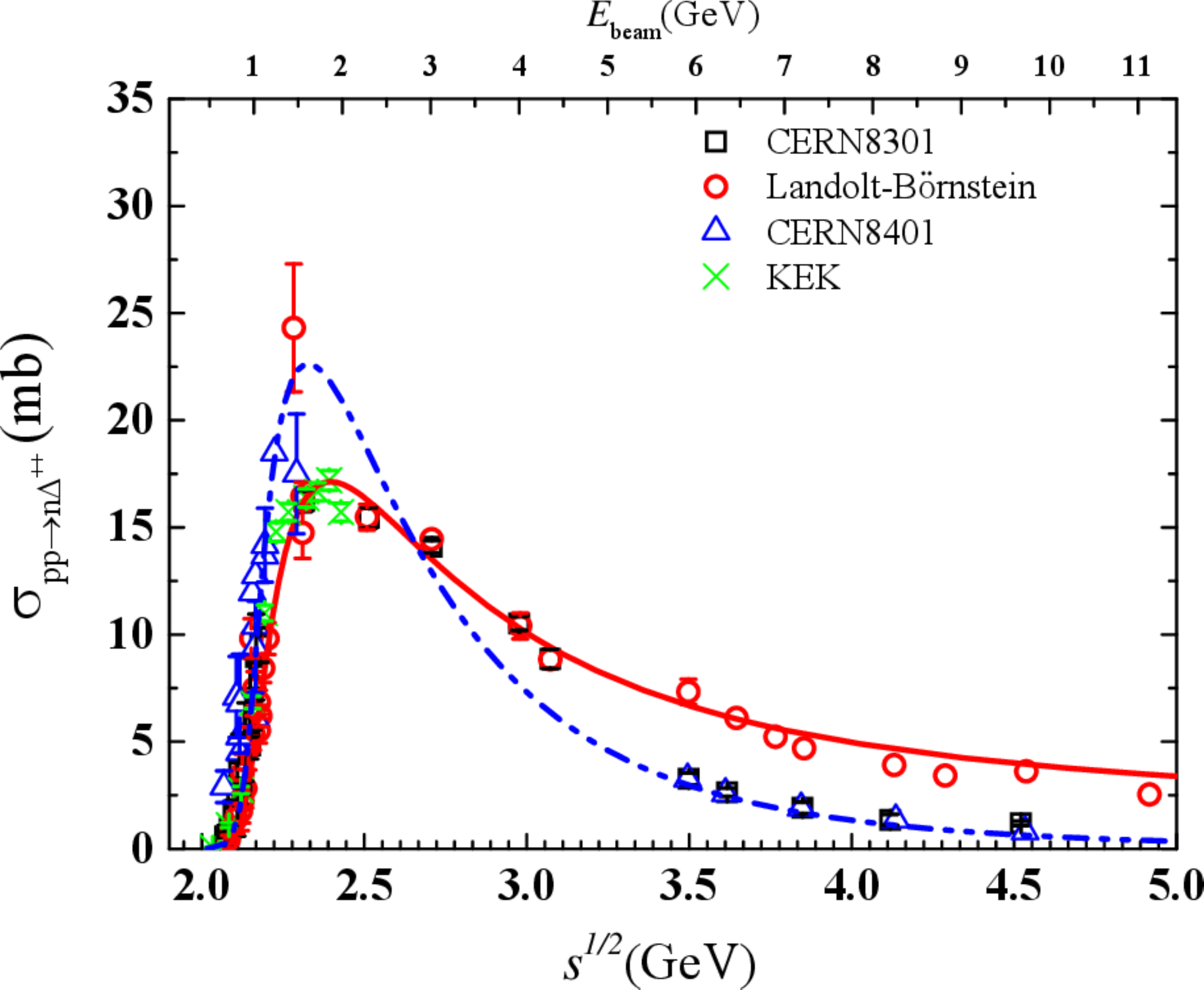}
    \caption{(Color online) $\sigma^*_{pp\rightarrow n\Delta^{++}}$ as a function of $s^{1/2}$ in free space, the experimental data from \cite{Baldini1987,Cern8301,Cern8401,KEK}. The blue dash line is for fitting the CERN8401\cite{Cern8401} and red dot line is for Landolt-B\"{o}rnstein  data \cite{Baldini1987}, respectively.}\label{figexp}
\end{center}
\end{figure}

Now, let's firstly check the mass dependence of the extracted M-matrix, $\overline{|\mathcal{M}(m_{\Delta})|^2}$  in free space based on the OBEM. The details of M-matrix can be found in our previous paper\cite{Cui2018}, and the parameters in the expression of $|\mathcal{M}|^2$ are determined by fitting the measured cross section of $pp\to n\Delta^{++}$ \cite{Baldini1987}. Up to now, there are several groups published the measured cross section of $NN\to N\Delta$\cite{Cern8401,Cern8301,Baldini1987,KEK}. As shown in Fig.~\ref{figexp}, the measured cross section of $pp\to n\Delta^{++}$ still have 3-5 mb uncertainties around $\sqrt s \sim$ 2.2 GeV and above 3.0 GeV. Two typical values of cross section of $pp\to n\Delta^{++}$, CERN8401(blue triangles) \cite{Cern8401} and Landolt-B\"{o}rnstein\cite{Baldini1987} (red circles), are chosen to adjust the parameters in the M-matrix since they are two extreme case of the published data of $pp\to n\Delta^{++}$. 
%

\begin{figure}[htbp]
\begin{center}
\includegraphics[scale=0.36]{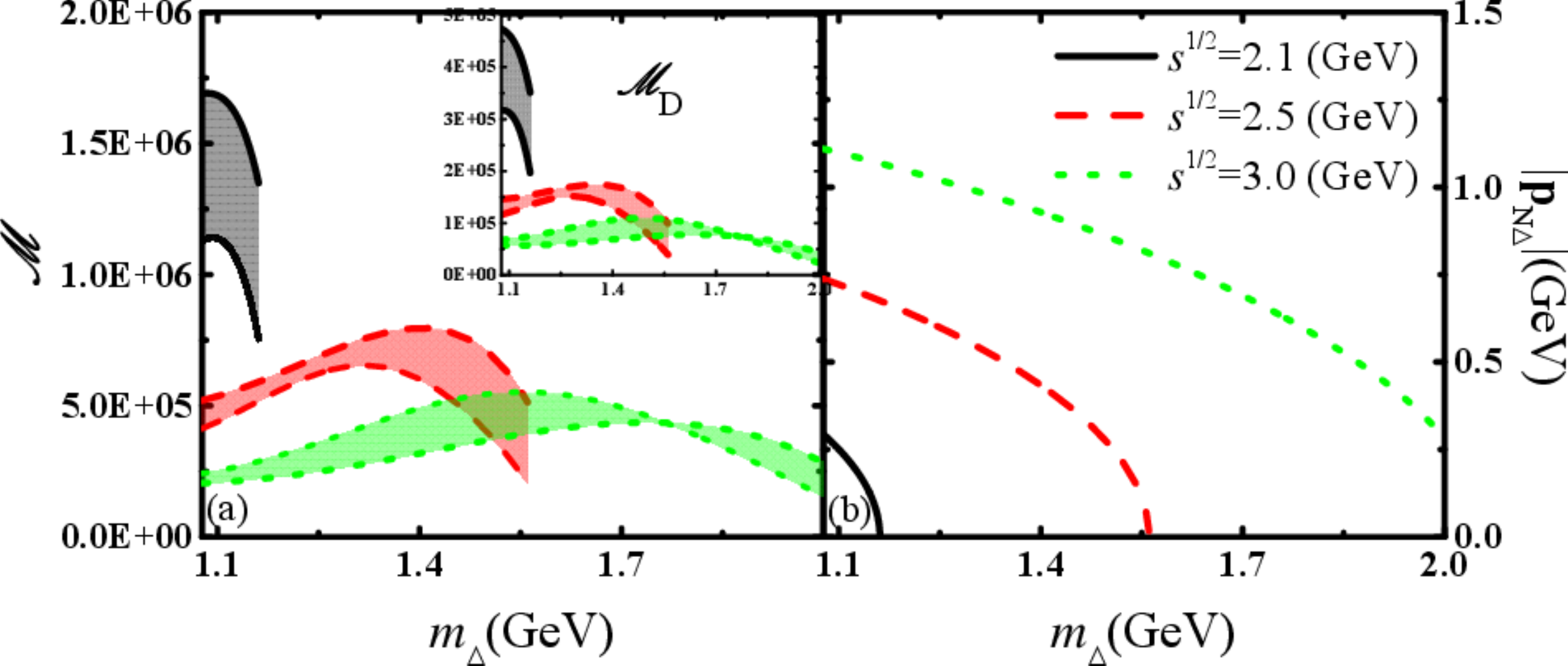}
\caption{(Color online) (a) $\mathscr{M}=\frac{1}{I^2_i}\int \sum_{s}|\mathcal{M}|^2 d\Omega$ as a function of $m_{\Delta}$  in free space for the  energy $s^{1/2}$ are 2.1, 2.5 and 3 GeV respectively, and the insert figure is $\frac{1}{I^2_i}\int \sum_{s} |\mathcal{M}_{D}|^2 d\Omega$ (GeV$^{-4}$) in Ref. \cite{Danielewicz1991}. (b)  $|\textbf{p}_{N\Delta}|$ as a function of $m_{\Delta}$  in free space .}\label{fig3v2}
\end{center}
\end{figure}

In the Fig.~\ref{fig3v2} (a), we plot the angular integrated isospin independent M-matrix as a function of $m_{\Delta}$ at the total energy $s^{1/2}=$ 2.1, 2.5 and 3.0 GeV respectively. The shadow region corresponds to the M-matrix with their experimental uncertainties which are obtained with experimental data from CERN8401\cite{Cern8401} and Landolt-B\"{o}rnstein data\cite{Baldini1987}. The range of $m_\Delta$ is from $m_N+m_\pi$ to $\sqrt{s}-m_N$, where the maximum value of $m_\Delta$ depends on the energy in the process of $NN\to N\Delta$. The isospin independent M-matrix is obtained by normalized the M-matrix with their isospin factors, i.e.,
\begin{equation}
\mathscr{M}=\frac{1}{I^2_i} \int \sum_{s}|\mathcal{M}|^2 d\Omega=\frac{(2s_1+1)(2s_2+1)}{I^2_i} \int \overline{|\mathcal{M}|^2} d\Omega.
\label{eq:Mint}
\end{equation}
$I_{i=d,e}$ is the isospin factor as same as in Ref.\cite{Cui2018,Cui2019}, and $I^2_{d,e}(n\Delta^{++}\to pp) =I^2_{d,e}(p\Delta^{-}\to nn)=2$ and $I^2_{d,e}(\text{other channels})= 2/3$. As shown in left panel of Fig.~\ref{fig3v2}, the behaviors of of $\mathscr{M}$ as a function of $m_\Delta$ based on OBEM clearly shows that $\overline{|\mathcal{M}(m_{\Delta})|^2}$ depends on $m_\Delta$ whatever the experimental data one used. In order to understand the assumption of mass dependence of M-matrix mentioned in Danielewicz's method in Ref.\cite{Danielewicz1991}, we also present the $\mathscr{M}_{D}=\frac{1}{I_i^2}\int|\mathcal{M}_{D}|^2 d\Omega$ in the inset of Fig.~\ref{fig3v2} (a), which is
as same convention as in Ref.\cite{Danielewicz1991} with GeV$^{-4}$. At the energy range we selected, our calculations illustrate that $|\mathcal{M}_D|^2$ obviously depend on $m_\Delta$ in all the mass region where $\Delta$ can be produced. It can be understood from the formula of M-matrix as in Eq.(22) in Ref. \cite{Cui2019}. For example, if one analyze the power of $m_\Delta$ in the M-matrix, it will be roughly in the form with $m_\Delta^2$.
At higher energies, the $\Delta$ mass dependence of M-matrix becomes weak which means the assumption on the calculation of $N\Delta\to NN$ in reference\cite{Danielewicz1991} is reasonable.

Another point need to be investigated is the mass dependence of $|\textbf{p}_{N\Delta}(m_\Delta)|$ (here  $|\textbf{p}_{N\Delta}(m_\Delta)|$ is $|\textbf{p}_{34}(m_\Delta)|$) in Eq.~\ref{eq:ratio}. In Fig.~\ref{fig3v2} (b), we present the mass dependence of $|\textbf{p}_{N\Delta}(m_\Delta)|$ at different energies, where the $|\textbf{p}_{N\Delta}(m_\Delta)|$ decreases with the mass of $\Delta$ and the mass dependence is much sharper at the lower energies than that at higher energies. The panels in figure ~\ref{fig3v2} show that both $\overline{|\mathcal{M}(m_{\Delta})|^2}$ and $|\mathbf{p}_{N\Delta}(m_{\Delta})|$ in Eq.~\ref{eq:ratio} obviously depend on the mass of $\Delta$, especially at lower energies.

Clearly, Fig.~\ref{fig3v2} tells us that the $\Delta$ mass dependence (which depends on the system energy) of M-matrix and $|\textbf{p}_{N\Delta}(m_\Delta)|$ are not ignorable, which can influence the accuracy of calculations of the  $N\Delta\to NN$ cross section based on the detailed balance by using Eq.~(\ref{eq:dbpawel}) or Eq.~(\ref{eq:db92}). We select three typical values of $m_\Delta$ to understand the precision of the different ways to estimate the $\sigma_{N\Delta \to NN}$. The minimum mass of $\Delta$ ($m_{\Delta}=m_{\text{min},\Delta}=1.077$ GeV), pole mass ($m_{0,\Delta}=1.232$ GeV), and $m_\Delta=1.387$ GeV which corresponds to the maximum mass of $\Delta$ production in heavy ion collisions at the beam energy of 1 GeV. 
Since others data also give the similar $m_{\Delta}$ dependence of M-matrix as shown in Figure~\ref{fig3v2}, in the following, we use the M-matrix with their parameters are extracted based on the data from Landolt-B\"{o}rnstein \cite{Baldini1987} to valuate the accuracy and validity of the method to calculate the cross section of $N\Delta\to NN$.

\begin{figure}[htbp]
\begin{center}
    \includegraphics[scale=0.3]{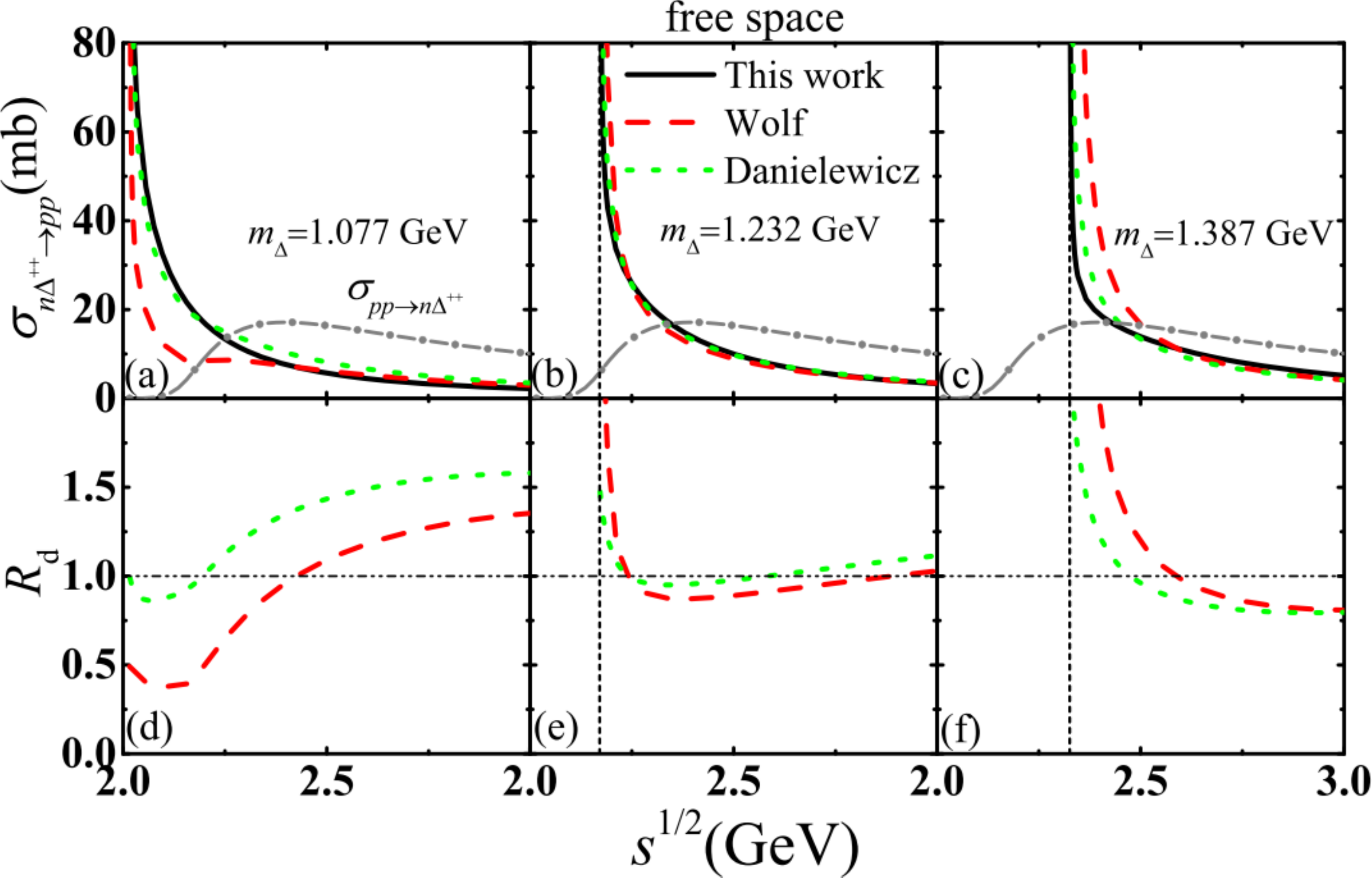}
    \caption{(Color online)  The upper panel is $\sigma_{n\Delta^{++}\rightarrow pp }$  as a function of $s^{1/2}$   for different types of the detailed balance  at $m_{\Delta}=1.077$ GeV, $1.232$ GeV and 1.387 GeV in free space. The bottom panel is $R_{d}$ as a function for different detailed balance.}\label{fig3}
\end{center}
\end{figure}

Fig.~\ref{fig3} (a)-(c) present the results of $\sigma_{n\Delta^{++} \rightarrow pp }$ as a function of $s^{1/2}$ in free space at $m_\Delta=1.077$, $1.232$ and $1.387$ GeV. The black solid lines are the $\sigma^{th}_{n\Delta^{++} \rightarrow pp}$ which are directly calculated from the M-matrix element of $N\Delta\to NN$ based on the scattering theory. This result is a benchmark for evaluating other approaches for calculation of the cross section of $n\Delta^{++} \rightarrow pp$.  The red dashed lines are the results obtained with the method adopted in Wolf's  work \cite{Wolf1992,Wolf1993,Engel1994}, i.e. Eq.(\ref{eq:db92}), without considering the mass dependence of $|\textbf{p}_{N\Delta}|$, and we named it as $\sigma^{DB,W}_{n\Delta^{++} \rightarrow pp}$. The green dotted lines are results obtained from the method proposed by Danielewicz, i.e. Eq.(\ref{eq:dbpawel}), in which the mass dependence of $|\mathcal{M}|^2$ is neglected, and we named it as $\sigma^{DB,D}_{n\Delta^{++} \rightarrow pp}$.
All the methods predict that there is large $\Delta$ absorption cross section around the threshold of $n\Delta^{++} \rightarrow pp$ process which increases with the $m_\Delta$ increasing, and $\sigma_{n\Delta^{++} \rightarrow pp}$ decreases with the energy increasing. However, both methods can not well reproduce the $\sigma_{n\Delta^{++} \rightarrow pp}$ around the threshold energy if the mass of $\Delta$ is away from the pole mass, $m_\Delta=m_{0,\Delta}$=1.232 GeV. Other channels of $N\Delta\to NN$ have similar results since the differences only come from the isospin factor.

To clearly see the deviations, we present the ratio which is defined as $R_{d}=\sigma_{\Delta N\to NN}^{DB}/\sigma_{\Delta N\to NN}^{th}$ in the Fig.~\ref{fig3} (d)-(f) for different mass of $\Delta$. The $R_d$ has the same values for all the channels of $N\Delta\to NN$ since the contributions from isospin factor are cancelled in the ratio. $R_d=1$ means the cross section of $N\Delta\to NN$ is described by the proposed method. Red dashed lines are the results of $\sigma^{DB,W}_{\Delta N\to NN}/\sigma^{th}_{\Delta N\to NN}$, and green dotted lines are the results of $\sigma^{DB,D}_{\Delta N\to NN}/\sigma^{th}_{\Delta N\to NN}$. For $m_{\Delta}=m_{0,\Delta}$, both methods can well reproduce the theoretical values of $\sigma^{th}_{N\Delta\to NN}$ except for the $s^{1/2}<2.2$ GeV, where the Danielewicz's method is much closer to the theoretical one compared to Wolf's method. If $m_\Delta=1.076$ or $1.387$ GeV, larger deviations can be found near the threshold energy of $N\Delta\to NN$ process (close to vertical dashed lines). For example, if $m_\Delta$ is close to the minimum mass of $\Delta$, both methods in Ref.\cite{Danielewicz1991,Wolf1992} underestimate the $\Delta$ absorption cross section at $s^{1/2}<2.2$ GeV, and the deviation is less than 20\% for Danielewicz method and larger than 50\% for Wolf's approach. Both methods overestimate the $\Delta$ absorption cross section and the deviation is close to 50\% for Danielewicz method while Wolf's approach gives the deviation less 40\% at $s^{1/2}>2.5$ GeV. At large mass region of $\Delta$, both methods overestimate the $\Delta$ absorption cross section at $s^{1/2}<2.47$ GeV, but they underestimate the $\Delta$ absorption cross section at $s^{1/2}>2.6$ GeV. The above comparison suggests that the mass dependence of M-matrix as well as $|\textbf{p}_{N\Delta}(m_\Delta)|$ should be taken into account for precise calculation of $NN\to N\Delta$ cross section. 

\begin{figure}[htbp]
\begin{center}
\includegraphics[scale=0.37]{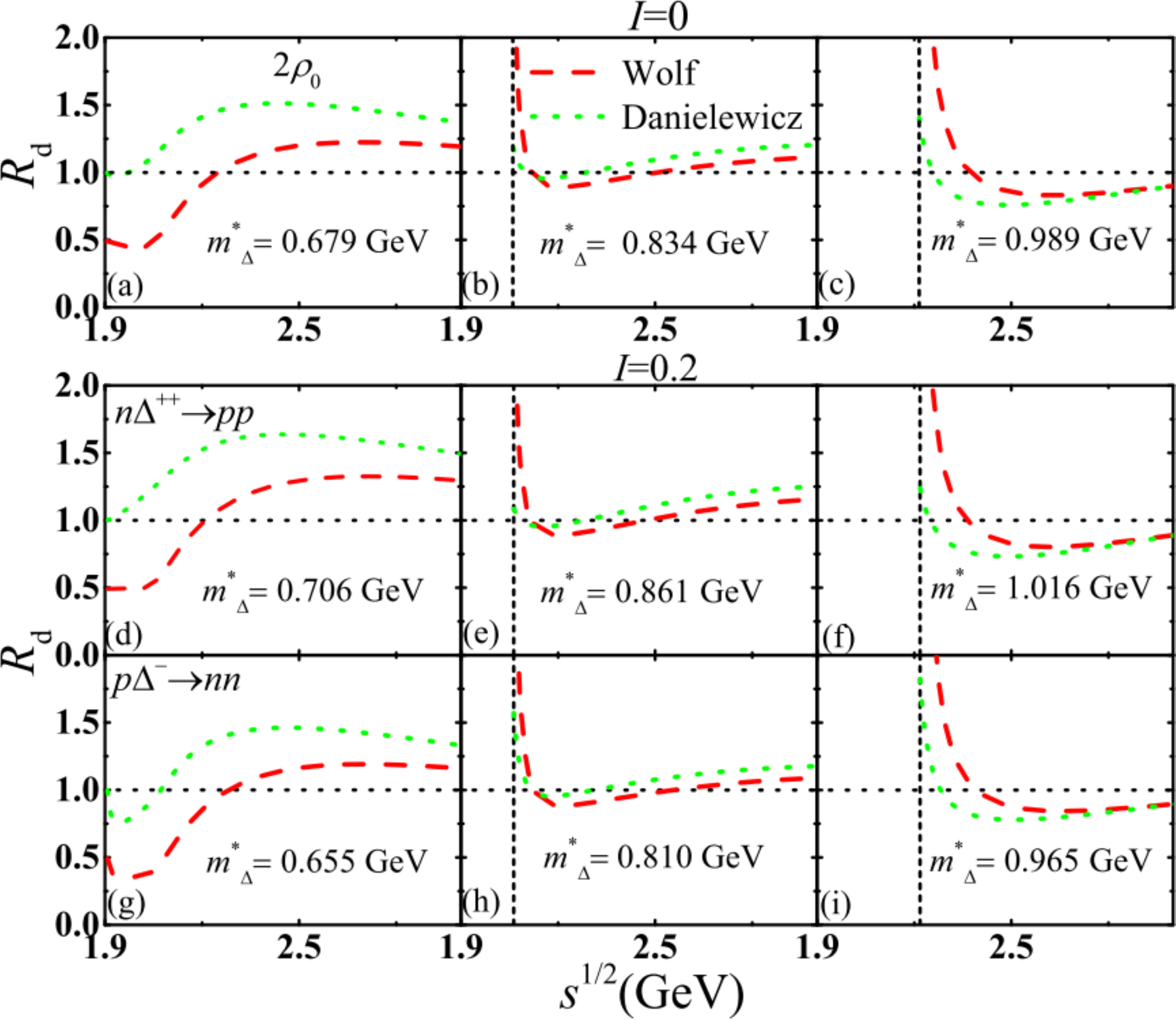}
\caption{(Color online) The upper panel is  in-medium $R_d$  as a function of $s^{1/2}$ for different types of the detailed balance  at $m^*_{\Delta}=m^*_{\Delta,\text{min}}$ (0.679 GeV), $m^*_{0,\Delta}$ (0.834 GeV) and $m^*_{\Delta}$=0.989 GeV in symmetric nuclear matter ($I$=0) at $2\rho_0$. The bottom panel is $R_d$ for  $n\Delta^{++}\to pp$ and $p\Delta^{-}\to nn$ at $2 \rho_0$  for different $m^{*}_{\Delta}$ in  asymmetric nuclear matter (nuclear asymmetry is $I$=0.2).}\label{fig4}
\end{center}
\end{figure}

By using the isospin independent  M-matrix, i.e. $\mathscr{M}$ in Eq.~\ref{eq:Mint}, the cross section for $N\Delta\to NN$ can be expressed as,
\begin{eqnarray}
&&\sigma_{N\Delta(m_{\Delta})\to NN}=\\\nonumber
&&\frac{1}{64\pi^2s} \frac{|\textbf{p}_{\text{12}}|}{|\textbf{p}_{N\Delta}(m_\Delta)|}\frac{I^2_i\mathscr{M}}{(2s_3+1)(2s_4+1)}\frac{1}{1+\delta_{N_1N_2}}
\end{eqnarray}
The values of $\mathscr{M}$, which are obtained by fitting the Landolt-B\"{o}rnstein data, are in the supplementary file. For $n\Delta^{++}\to pp$ and $p\Delta^{-}\to nn$ channels, $I_i^2=2$, while for $n\Delta^+\to np$, $n\Delta^0\to nn$, $p\Delta^+\to pp$, $p\Delta^0\to np$ channel, $I_i^2=2/3$. Hence, $\sigma_{n\Delta^{++}\to pp}$ : $\sigma_{p\Delta^{-}\to nn}$ : $\sigma_{n\Delta^+\to np}$ : $\sigma_{p\Delta^0\to np}$: $\sigma_{n\Delta^0\to nn}$ : $\sigma_{p\Delta^+\to pp}$ is 3:3:2:2:1:1. Since the mass dependence of the M-matrix is considered, the mass of $\Delta$ in the process of $NN\to N\Delta$ should also be sampled by considering the mass dependence of $|\mathcal{M}|^2$. Correspondingly, the $\Delta$ mass should be sampled with the following form,
\begin{equation}
P(m_\Delta)=\frac{\int_{m_{N}+m_{\pi}}^{m_\Delta}|\textbf{p}_{N\Delta}(m'_\Delta)|\times I^2_i\mathscr{M}\times f(m'_\Delta)dm'_\Delta}{\int_{m_{N}+m_{\pi}}^{\sqrt s-m_N}|\textbf{p}_{N\Delta}(m_\Delta)|\times I^2_i\mathscr{M}\times f(m_\Delta) dm_\Delta}
\end{equation}
Since the in-medium cross sections are adopted in the simulation of heavy ion collisions, it naturally requires us to check the accuracy of the calculations of the in-medium cross section of $N\Delta\to NN$ by using the methods proposed in \cite{Danielewicz1991,Wolf1992}.
Three $m^*_\Delta$ values are chosen to performance the $R_d$ as a function of $s^{1/2}$ at two times normal density ($2\rho_0$) as shown in Fig.~\ref{fig4}. The selected three $m^*_\Delta$ are similar to that in free space, but the $m_\Delta^*$ values also depend on the density, isospin asymmetry, and the charge state of $\Delta$. The upper three panels are for symmetric nuclear medium, and lower six panels are for isospin asymmetric nuclear medium. Similar to the results in free space, there are larger deviations at low or high $\Delta$ mass region than that around the pole mass regions, and $R_d$ values depend on the $m^*_\Delta$, $s^{1/2}$ and the channel of $N\Delta\to NN$ (for example, Fig.~\ref{fig4} (d) and (g)). 


In summary, we have valuated the methods to calculate $\sigma_{N\Delta\rightarrow NN}$ from $\sigma_{NN \rightarrow N\Delta}$ within the framework of OBEM. By comparing $\sigma_{N\Delta\rightarrow NN}$ from approximative methods to $\sigma^{th}_{N\Delta\rightarrow NN}$, which is
the exact calculation result from the M-matrix within the OBEM, our calculations show that both methods in Ref.\cite{Danielewicz1991} and \cite{Wolf1993} underestimate the low mass $\Delta$ absorption cross section and overestimate the large mass $\Delta$ absorption cross section near the threshold. We find that mass dependence of M-matrix should be considered, especially around the threshold energy. 
Considering the importance of mass dependence of M-matrix, we provide the supplementary data files for the $\mathscr{M}$ calculated by Eq.(\ref{eq:Mint}) which are fitted by the experimental data from Landolt-B\"{o}rnstein and CERN8401.

The influence of accurate calculations of $\sigma_{N\Delta\to NN}$ on heavy ion collisions by means of the transport model near the threshold energy is also worth to be investigated in the nearly future, because most of the $\Delta$ resonances participating in the process $N\Delta\to NN$ are low mass $\Delta$s. Another future interesting work is to valuate the calculation of other resonances with a much broader resonance width, it could be useful for deep understanding of the mechanism of particle productions in high energy heavy ion collisions.


\acknowledgments
This work has been supported by National Key R\&D Program of China under Grant No. 2018 YFA0404404, and National Natural Science Foundation of China under Grants No. 11875323, No. 11875125, No. 11475262, No. 11365004, No. 11375062, No. 11790323,11790324, and No. 11790325.







\end{document}